\documentclass[prb,aps,twocolumn,showpacs]{revtex4}

\usepackage{epsfig}

\begin{document}

\title{$^{23}$Na and $^{75}$As NMR Study of the Antiferromagnetism and the Spin Fluctuations on NaFeAs Single Crystals}

\author{L. Ma ${^1}$}
\author{G. F. Chen $^{1} $}
\author{Dao-Xin Yao ${^2}$}
\author{J. Zhang $^{3}$}
\author{S. Zhang ${^1}$}
\author{T.-L. Xia ${^1}$}
\author{Weiqiang Yu $^{1}$}
\email{wqyu_phy@ruc.edu.cn}

\affiliation{
$^1$Department of Physics, Renmin University of China, Beijing 100872, China\\
$^2$School of Physics and Engineering, Sun Yat-sen University, Guangzhou 510275, China\\
$^3$School of Energy and Power Engineering, North China Electric Power University, Beijing 102206, China\\
}
\date{\today}

\begin{abstract}

We report the $^{23}$Na and $^{75}$As NMR studies on NaFeAs single crystals. The structure transition temperature $T_S$ ($55$ K) and the spin density %%@
wave (SDW) transition temperature $T_{SDW}$ ($40.5$ K) are determined by the NMR line splits. The spin-lattice relaxation rates indicate that the spin %%@
fluctuations are strongly enhanced just below $T_S$ and drive a second order SDW transition. A fluctuating feature of the SDW ordering is also seen %%@
below the $T_{SDW}$. We further performed high-pressure NMR studies on NaFeAs, and found that the $T_{SDW}$ increases by $\sim$7 K and the magnetic %%@
moment increases by $\sim$$30\%$ under 2.5 GPa pressure. 
 
\end{abstract}

\maketitle

The discovery of superconductivity in iron pnictides \cite{Hosono_Jacs_130_3296} has attracted intense research interests, and so far high-temperature %%@
superconductivity is achieved in many iron-based compounds upon doping. In particular, three classes with similar structures, including the 1111 %%@
structure $R$FeAsO$_{1-x}$F$_x$ ($R$=La, Nd, Sm {\it etc.}) \cite{Chen_Nature_453_761,Chen_PRL_100_247002,Ren_MRI_12_105}, the 122 structure %%@
Ba(Fe$_{1-x}$Co$_x$)$_2$As$_2$/Ba$_{1-x}$K$_x$Fe$_2$As$_2$ \cite{Huang_prl}, and the 111 structure LiFeAs/NaFeAs \cite{Jin_CQ_LiFeAs, Chu_PRB, %%@
Clarke_LiFeAs} have been extensively studied. In their parent compounds, long-range antiferromagnetism (AFM), or the spin-density-wave (SDW), has been %%@
reported with a stripe-like magnetic structure \cite{Dai_Nature_453_899}. There are renewed concerns on the nature of the magnetism, regarding to %%@
whether the SDW follows a local moment or a Fermi surface nesting picture. For the 111 class, the magnetism appears to be very weak. In LiFeAs, %%@
superconductivity is observed instead of the SDW ordering \cite{Jin_CQ_LiFeAs}, although the SDW fluctuations are seen above $T_{C}$ \cite{Ma_prb}. In %%@
NaFeAs, the SDW order is observed with a low transition temperature \cite{Chen_PRL_102}. Its magnetic moment is reported to be 0.09 $\mu _B/$Fe by %%@
neutron scattering \cite{LiSL_NaFeAs}, in contrast to the larger values of about 0.4 $\mu _B/$Fe in the 1111 and about 1 $\mu _B/$Fe in the 122 parent %%@
compounds \cite{Dai_Nature_453_899, Huang_prl, Kaneko_Sr_NS}. From local density approximation (LDA) calculations, however, the band structures of all %%@
three classes are similar\cite{Jishi_ACM, Singh_PRB_78_094511, Zhang_PRB81_094505}. 

One important fact is that the SDW order only develops at or below the structure transition, namely the high-temperature tetragonal (HTT) to the %%@
low-temperature orthorhombic (LTO) transition. It is conjectured that the structure transition is important for the SDW ordering, and may also be %%@
important for the superconductivity. For instance, it has been argued that both the structure phase transition and the SDW ordering are driven by a %%@
ferro-orbital ordering caused by the $d_{xz}$ and $d_{yz}$ orbitals \cite{weiku_prl_09,kruger09_prb}. The structure transition $T_{S}$ and the SDW %%@
transition $T_{SDW}$ are well separated in NaFeAs and the 1111 class, while in the 122 class the two transitions occur simultaneously. These %%@
distinctive properties open a sight for studying the relation between the structure and the magnetism.

In this paper, we present our $^{23}$Na and $^{75}$As NMR studies on nominally undoped NaFeAs single crystals, mainly focusing on the interplay of the %%@
structure and the magnetism. First, we determined the sharp SDW transition temperature ($T_{SDW}\approx 40.5$ K) and the structure transition %%@
temperature ($T_S\approx 55$ K) directly from the NMR, and a commensurate magnetic moment of $0.32\pm 0.02\mu _B/$Fe far below $T_{SDW}$. Second, we %%@
study the correlation between the magnetism between the structure. Spin fluctuations, indicated by the spin-lattice relaxation rate, change behaviors %%@
across the structure transition. We also found that both the SDW transition temperature and the magnetic moment are enhanced significantly under %%@
pressure.

The NaFeAs single crystals were synthesized by flux-growth method with NaAs as flux, and the detailed growth procedure was reported elsewhere %%@
\cite{Chen_PRL_101_057007}. For NMR, we chose crystals with typical dimensions of 3$\times$2$\times$0.1mm$^3$. The crystals were characterized by the %%@
magnetization measurements. We performed $^{23}$Na  and $^{75}$As ( both with $S=3/2$) NMR studies, with the magnetic field either along the %%@
$ab-$plane or the $c-$axis. It is known that the chemical non-stoichiometry in the 111 class affects the magnetism significantly \cite{Ma_prb}. In %%@
this paper, we primarily report results of single crystals with very small superconducting volume ratio (less than $2\%$), which turn out to have very %%@
narrow NMR linewidth at both above and below the $T_{SDW}$, indicating good chemical stoichiometry. The SLRR is deduced from the spin recovery after %%@
an inversion pulse. For the high-pressure study, we used a 2.5 GPa pressure cell. Silicon oil is used as the pressure medium, and a piece of lead %%@
inside the NMR coil is used as a low-temperature manometer.

\begin{figure}
\includegraphics[width=8cm, height=6cm]{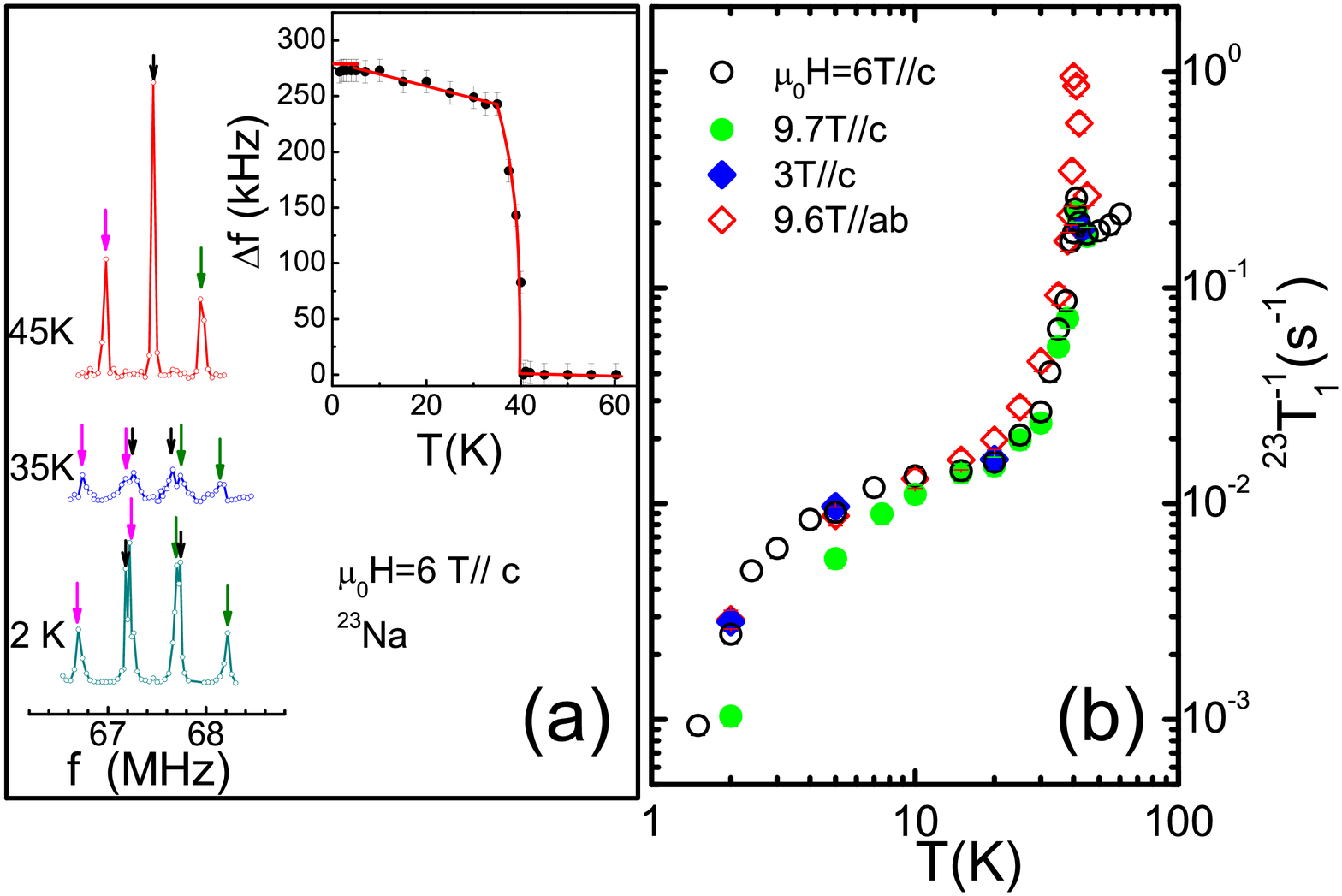}
\caption{\label{sdw}(color online) (a) The $^{23}$Na NMR spectra (one center transition and two satellites with $^{23}\nu _q \approx$0.525 MHz) at %%@
above and below the $T_{SDW}$ (40.5 K), with the magnetic field applied along the c-axis; Inset: The frequency shift of the $^{23}$Na center %%@
transition below the SDW transition; (b) The spin-lattice relaxation rate of $^{23}$Na with different field orientations and field amplitudes.}
\end{figure}

\begin{figure}
\includegraphics[width=7cm, height=6cm]{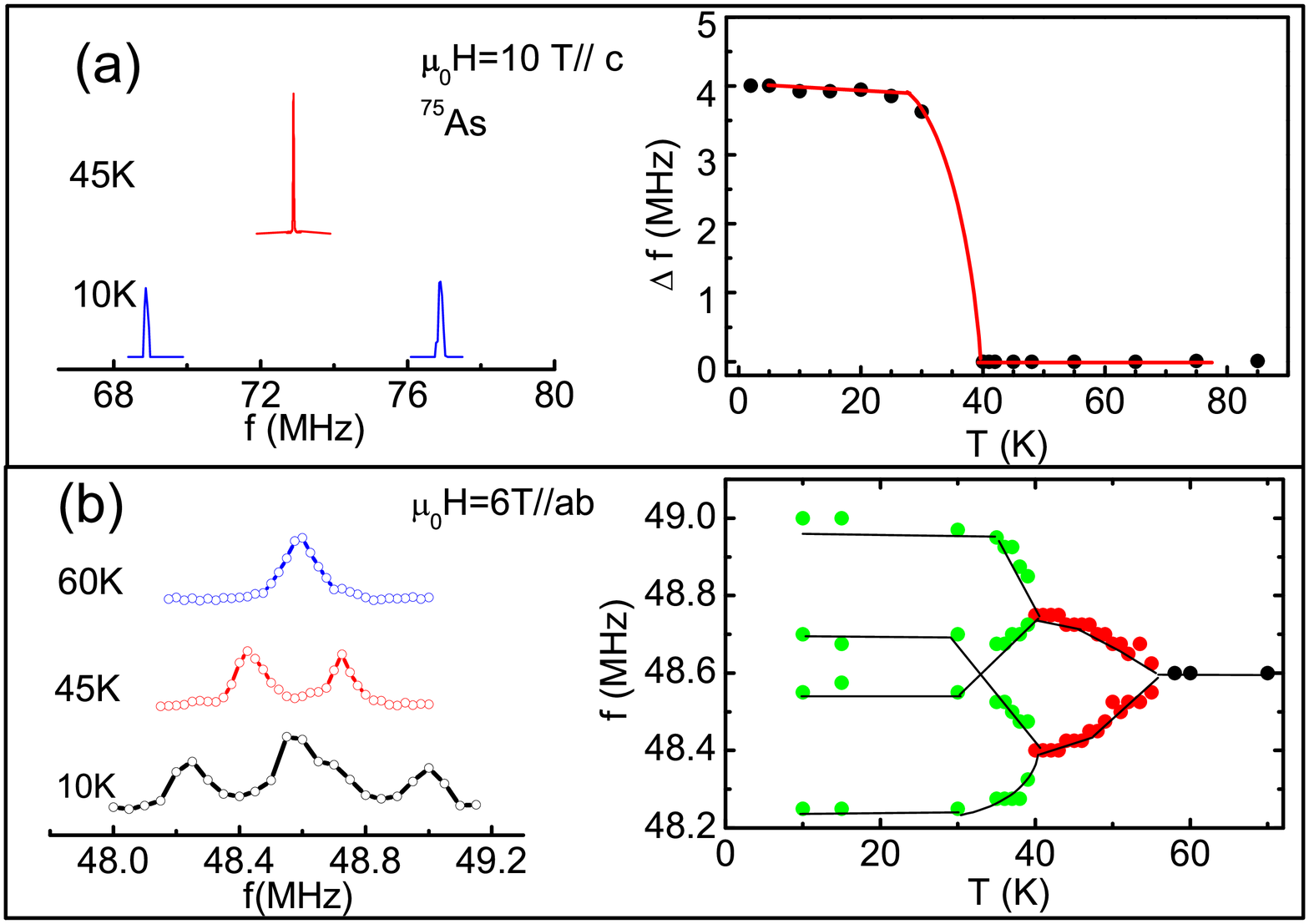}
\caption{\label{twin}(color online) (a) The $^{23}$As center transition with the magnetic field along the c-axis (satellites not shown). Inset: The %%@
frequency shift of the center transition; (b) The $^{75}$As NMR satellite spectrum ($^{75}\nu _q \approx 9.5$ MHz) at different temperaures; Inset: %%@
The temperature dependence of one $^{75}$As NMR satellite frequency with the field applied primarily along one a-axis but $3^o$ off to the c-axis.  }
 \end{figure}

The SDW transition is seen clearly by the $^{23}$Na  spectra with field applied along the c-axis, as shown in Fig.~\ref{sdw}(a). The nuclear %%@
quadrupole splitting is $^{23}\nu _q \approx 0.525$ MHz, as evidenced by one center transition and two satellite lines at 45 K. As the sample is %%@
cooled below 40.5 K, each line splits into two species with equal frequency shift (denoted by the arrows with the same color), which shows the SDW %%@
ordering. The relative frequency split of the central transition, 2$\Delta f$, serves as an order parameter of the SDW transition. As shown in  %%@
Fig.~\ref{sdw} (a) inset, the transition width is very narrow, and $\Delta f$ almost saturates at 5 K below the $T_{SDW}$. $^{75}$As spectra also have %%@
a line splitting below the SDW transition with $H\parallel c$, as shown in Fig.~\ref{twin} (a). The increase of $^{75}$As frequency shift $\Delta f$ %%@
below the SDW transition is also shown in Fig.~\ref{twin}(a).

With $H\parallel ab$, the line splitting is absent for both $^{23}$Na and $^{75}$As (data not shown). The splitting of the NMR spectrum indicates two %%@
internal static hyperfine fields $H_{in}=\pm \Delta H$ along the c-axis. The observation of c-axis internal field on $^{75}$As suggests a stripe AFM, %%@
due to an off-diagonal hyperfine coupling between the $^{75}$As nuclei and the Fe moments\cite{Kita_JPSJ_77_114709}. The sharp NMR spectrum far below %%@
the $T_{SDW}$ suggests a commensurate order, which is consistent with the neutron data \cite{LiSL_NaFeAs}. It is reasonable to have the same direction %%@
of the internal field for $^{23}$Na and $^{75}$As, since two nuclei are on the inversion symmetry position to each other.

The HTT to the LTO structure transition can be inferred from $^{75}$As satellites. In Fig.~\ref{twin}(b), one $^{75}$As satellite (the high frequency %%@
one) is shown at different temperatures, with the magnetic field applied primarily along an $a$-axis. When the sample is cooled down, the spectrum %%@
splits into two species at about 55 K, a signature of the structure transition. The spectrum splitting is caused by sample twinning, and domains with %%@
field along the $a$-axis and $b$-axis give different resonance frequencies. When the temperature goes down, the spectrum further splits at the SDW %%@
transition, because the applied field is slightly off the $ab$ plane.  

We also analyze the SDW moment from spectrum splitting $2\Delta f$  of $^{75}$As and $^{23}$Na. In Fig.~\ref{sdw}(a) and Fig.~\ref{twin}(a), the %%@
values of $\Delta f$ for $^{75}$As and $^{23}$Na are shown at different temperatures. The frequency shift $\Delta f$ is about 4.0 MHz at $T=$2 K, %%@
corresponding to a hyperfine field $H_{in}\approx$ 0.55 T for $^{75}$As. For iron pnictides, the magnetic moment $m$ is known to follow $m=H_{in}/4 %%@
^{75}A_{hf}^{ac}$, where $^{75}A_{hf}^{ac}$ is the off-diagonal hyperfine coupling constant which is not directly measurable %%@
\cite{Kita_JPSJ_77_114709}. Here we employ $^{75}A_{hf}^{ac}\approx 0.43T/\mu _B$ from BaFe$_2$As$_2$\cite{Kita_JPSJ_77_114709}, and estimate the %%@
magnetic moment as $0.32\pm 0.02 \mu _B$/Fe, which is close to the 1111 class. Comparing with the frequency shift of $^{23}$Na, it gives %%@
$^{23}A_{hf}^{ac}\approx 0.027T/\mu _B$.

\begin{figure}
\includegraphics[width=7cm, height=5.5cm]{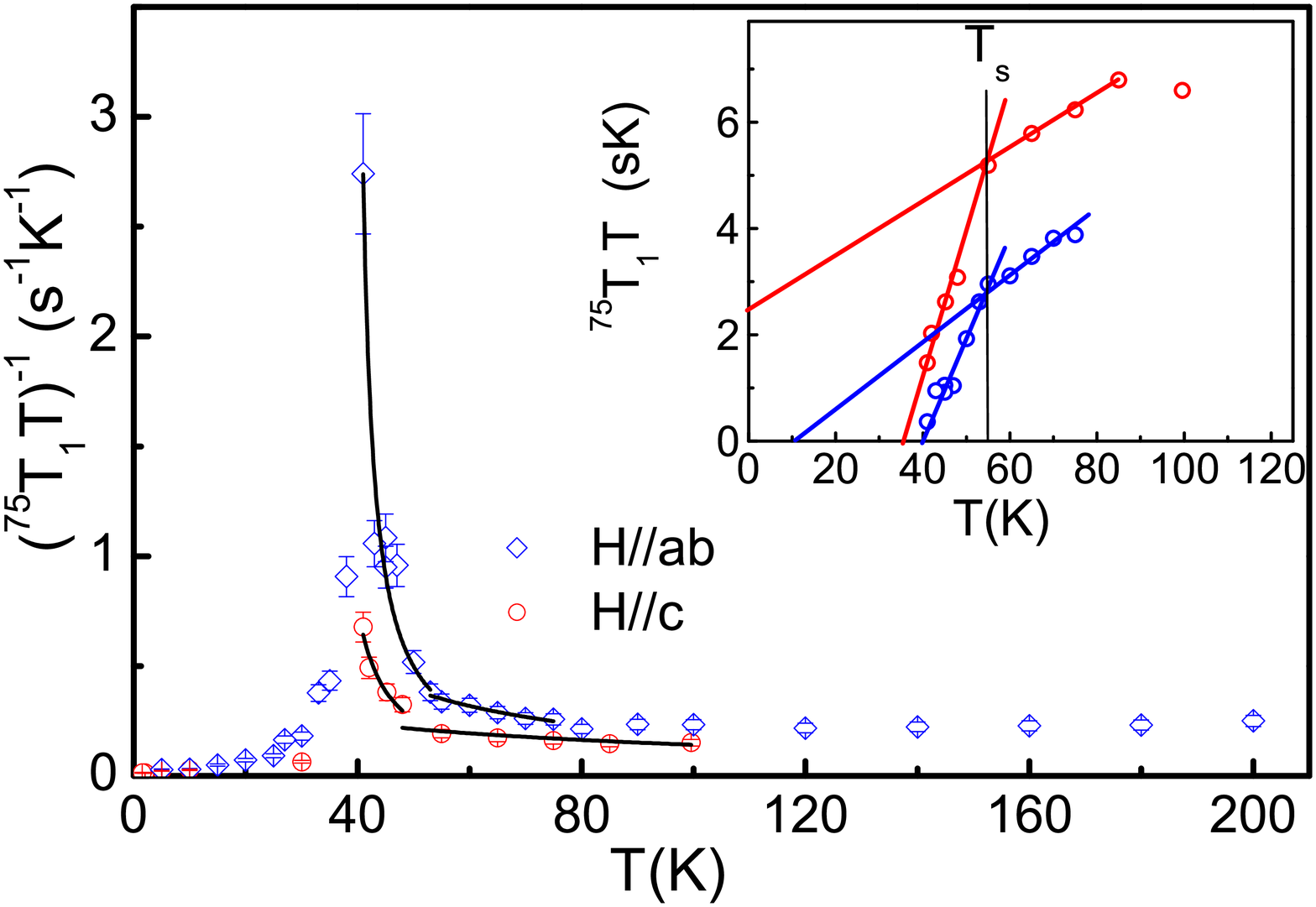}
\caption{\label{ast1hc}(color online) The temperature dependence of the $1/^{75}T_1T$ of a NaFeAs single crystal, with $\mu _0H=10 T$. Inset: The %%@
temperature dependence of $^{75}T_1T$, and the straight lines are fittings to $^{75}T_1T\sim (T+\Theta)$ for temperatures above and below the %%@
$T_{S}$.}
\end{figure}

Now we study the spin fluctuations from the spin lattice relaxation rate $1/T^{ab}_1$ ($H\parallel ab$) and $1/T^{c}_1$ ($H\parallel c$). The %%@
temperature dependence of $1/^{75}T^{ab}_1T$ and $1/^{75}T^c_1T$ are shown respectively in Fig.~\ref{ast1hc}. From room temperature to 100 K, the %%@
$1/^{75}T_1T$ decreases, which is also seen in other iron pnictides \cite{Ning_PRL_104}. Below 100K, the $1/T_1T$ increases as the temperature drops %%@
with an upturn like behavior, and the $1/^{75}T_1T$ is anisotropic with $^{c}T_1/^{ab}T_1\approx 1.5$. The upturn and the anisotropy of the SLRR are %%@
indications of strong SDW spin fluctuations \cite{Kita_JPSJ_77_114709}. The SDW transition temperature is shown at 40.5 K by the divergence of the %%@
$1/^{75}T_1T$. Below $T_{SDW}$, the $1/^{75}T_1T$ drops quickly because of the suppression of the spin fluctuations. Notably $1/^{75}T_1T$ saturates %%@
with a constant value 0.015 s$^{-1}$K$^{-1}$ at low temperatures, similar to that of BaFe$_2$As$_2$ \cite{Kita_JPSJ_77_114709}.

A correlation between the structure transition and SDW ordering can be seen from the SLRR data across the the structure transition. The spin %%@
fluctuations are significantly enhanced just below the structure transition, which is shown by the rapid increase of the $1/^{75}T_1T$ below 55K %%@
(Fig.~\ref{ast1hc}). Such behavior is better shown by the temperature dependence of $^{75}T_1T$ (Fig.~\ref{ast1hc} (inset)). The antiferromagnetic %%@
spin fluctuations are usually shown by a Curie-Weiss like behavior with $1/T_1T=A/(T+\Theta)$. Fitting $^{75}T^{ab}_1T$ (or $^{75}T^{c}_1T$) with %%@
$T_1T=(T+\Theta)/A$, two different lines are needed for above and below the structure transition. For $H\parallel ab$, the fitting gives $\Theta %%@
_{ab}\approx -10\pm 5K$ (above $T_{S}$) and $\Theta _{ab}\approx -40\pm 1$ K (below $T_{S}$). 

The above values of $\Theta _{ab}$ has two implications. First, above $T_{S}$, $\Theta _{ab}$ is negative, which suggests that the spin fluctuation %%@
could lead to the magnetic ordering at a finite temperature even without the structure transition. This supports the $J_1$-$J_2$-$J_c$ model that the %%@
structure transition is in fact caused by the spin fluctuations\cite{chen_prb08, sachdev_prb08}. Second, the value $\Theta _{ab}\approx -T_{SDW}$ %%@
below $T_{S}$ clearly indicates that the SDW is a second order transition. Since the large value of $-\Theta$ suggest stronger spin fluctuations %%@
\cite{Moriya, Ning_PRL_104}, the enhancement of the $-\Theta _{ab}$ below the structure transition and the second order SDW transition suggest that %%@
there is an interplay of the structure transition and the magnetism. In the LTO phase, the crystal $a$-axis and the $b$-axis are inequivalent, which %%@
gives an anisotropic coupling of $J_{1a}$ and $J_{1b}$ and may help the formation of the SDW ordering. A recent ARPES study on NaFeAs which revealed a %%@
band shift below $T_s$\cite{Feng_DL_NaFeAs}, which may suggest that the band structure is also involved in the magnetic ordering.

Furthermore, since the structure phase transition and the magnetic phase transition are separated, the interlayer coupling $J_c$ in NaFeAs is probably %%@
weaker than the 122 compounds from the picture of magnetically driven structure phase transition. From the measured values, we have %%@
$(T_S-T_{SDW})/T_{SDW}=0.358$.  According to Ref.\cite{chen_prb08}, we can deduce that the interlayer coupling in NaFeAs is also weaker than the 1111 %%@
class. The weak interlayer coupling lifts the degeneracy of the structure phase transition and the SDW ordering. In this picture, the SDW transition %%@
is naturally a second order type.

\begin{figure}
\includegraphics[width=8cm, height=8cm]{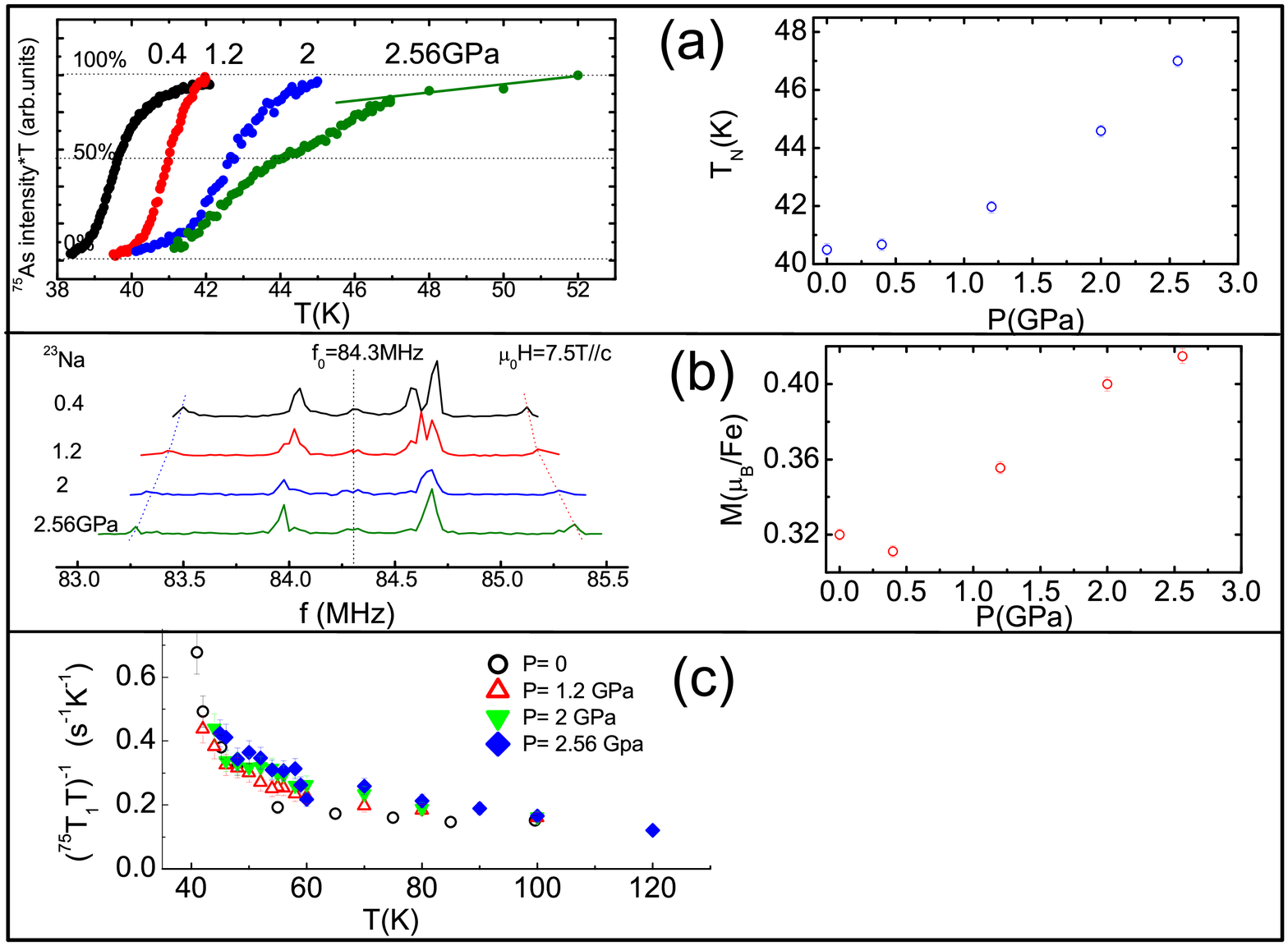}
\caption{\label{pressure}(color online) (a): The temperature dependence of the $^{75}$As paramagnetic spectral weight at different pressures. The %%@
sharp drop of the signal indicates the onset of the SDW ordering. Inset: The onset $T_{SDW}$ determined from the sudden drop of the spectral weight at %%@
different pressures. (b): The $^{23}$Na NMR spectrum of NaFeAs under different pressures ($T=2$ K). Inset: The SDW moment of NaFeAs estimated from the %%@
line splitting of $^{23}$Na at different pressures. (c): The $^{75}$As spin-lattice relaxation rate at different pressures.}
\end{figure}    

We further study the structure effect to the magnetism under high pressure up to 2.56 GPa, and find that spin fluctuations, the $T_{SDW}$ and the %%@
moment size are greatly enhanced by pressure. The increase of the spin fluctuations under pressure is seen by the increase of $1/^{75}T^c_1T$ %%@
(Fig.~\ref{pressure} (c)). In Fig.~\ref{pressure} (a), the normalized spectral weight of $^{75}$As in the paramagnetic phase is shown at different %%@
temperatures. The sudden drop of the spectral weight upon cooling indicates the onset of the SDW ordering (defined as $T_{SDW}$). As shown in %%@
Fig.~\ref{pressure} (a) (inset), the $T_{SDW}$ increases with pressure, from 40.5 K at the ambient pressure to 47 K at 2.56 GPa. The magnetic moment %%@
also seems to increase. In Fig.~\ref{pressure}(b), the spectrum of $^{23}$Na is shown at different pressures ($T=2$ K). The spectrum splitting clearly %%@
increases with pressure.  The magnetic moment increases from 0.32$\mu _B$ at $P=0$ to 0.41$\mu _B$ at $P=2.56$ GPa. Our estimation is much larger than %%@
the neutron scattering data\cite{LiSL_NaFeAs}. We note our estimation employs the hyperfine coupling constant obtained from BaFe$_2$As$_2$, which can %%@
be off if the hyperfine coupling constant is very different. The difference could also be partly caused by chemical stoichiometry %%@
\cite{Chen_PRL_101_057007, Jin_CQ_LiFeAs}, since NMR is a local probe and the neutron scattering measures the averaged magnetic moment.

The high-pressure data may suggest that the interlayer coupling $J_c$ plays a crucial role in determining the phase transitions and the magnetic %%@
moment. If we assume the lattice spacing along c-axis changes more significantly with pressure than  the $a$- or the $b$-axis,  the pressure is %%@
probably more effective in strengthening the interlayer coupling $J_c$ to enhance the SDW ordering \cite{yaoerica10}. Under pressure, the interlayer %%@
coupling $J_c$, mainly from $d_{xz}$ and $d_{yz}$ orbits, increases because the lattice constant becomes smaller. Unfortunately, our NMR satellite is %%@
broadened at higher pressures, and we cannot determine the structure transition. Further study is necessary to disclose the relation between the %%@
$T_{SDW}$ and $T_{S}$ under pressure. Nevertheless, the increased SDW moment (by $30\%$ at 2.56 GPa) and the $T_{SDW}$ (by $18\%$ at 2.56 GPa) draw %%@
the NaFeAs high-pressure phase closer to the 1111 class, which again supports that the c-axis coupling is weaker in NaFeAs.

We also observe a fluctuating feature of the SDW ordering below the $T_{SDW}$, which is not well understood. First, the spectrum splitting $\Delta f$ %%@
of $^{23}$Na (see Fig.~\ref{sdw}(a) inset) shows two steps. The $\Delta f$ increases quickly below $T_{SDW}$, and then increase slowly below 35K, %%@
followed by a full saturation at $T=10$ K. Second, $^{23}$Na linewidth broadens significantly between 40 K and 30 K (see Fig.~\ref{sdw} (a)), and %%@
narrows again at low temperatures. For $^{75}$As, the spectrum is not measurable between 40 K and 30 K, which indicates that the broadening is more %%@
significant. These temperature behavior suggest the the SDW ordering still fluctuates in an intermediate temperature range below the $T_{SDW}$, %%@
unlikely due to an disorder effect. 

Thermally activated domain walls may be an explanation of the above observation. In iron pnictides, the spin frustration is strong, and several types %%@
of domain walls may be thermally activated in the stripe phase, such as the antiphase type and/or the c-axis mis-alignment boundaries %%@
\cite{Mazin_nphys, Curro_NJP}. Under such circumstances, the NMR spectrum narrows and the magnetic moment saturates simultaneously with decreasing %%@
temperature. Another possible explanation is that the fluctuating features are caused by an incommensurate modulation on the SDW, which becomes %%@
commensurate again at low temperatures.

In summary, our data reveal the interplay between the lattice structure and the magnetic ordering in the undoped NaFeAs. First, our negative value of %%@
$\Theta _{ab}$ above the structure transition, from the fitting $1/T_1T=A/(T+\Theta)$, is consistent with the proposal of the magnetically driven %%@
structure phase transition. Second, the increase of $-\Theta _{ab}$ and the second order SDW transition below the structure transition suggest that %%@
the lattice/band structure in return strengthens the magnetic ordering. Third, the SDW ordering of NaFeAs is greatly enhanced upon pressure. We also %%@
observe the fluctuating features of the SDW ordering in an intermediate temperature below $T_{SDW}$, which may be caused by thermal fluctuations of %%@
domain walls, and/or high-temperature incommensurability. These magnetic properties should be important inputs for constructing the microscopic model %%@
of magnetism in iron pnictides. 
  
The Authors acknowledge W. Bao, S. E. Brown, P. Dai, X. Dai, T. Li, Z. Lu, F. Ning, B. Normand, X. Wang, and T. S. Zhao for helpful discussions. This %%@
work is supported by the NSFC (Grant Nos. 11074304, 10974254, and 11074310) and the National Basic Research Program of China (Contract Nos. %%@
2010CB923000 and 2011CBA00100).  

%\bibliography{NFA_ref}

\end{document}